\documentclass[11pt,a4paper]{article}
\usepackage{amsmath}
\usepackage{enumerate}
\usepackage{amssymb}
\usepackage{graphicx}
\usepackage{multirow}
\usepackage{xcolor}
\usepackage{tablefootnote}
\usepackage{threeparttable}
\usepackage{bm}
\usepackage[normalem]{ulem}
\usepackage{physics}
\usepackage{cancel}
\usepackage{soul}

\usepackage{cite} 
\usepackage{amsmath,mathtools}
\usepackage{float}
\usepackage{subcaption} 
\usepackage{comment}
\usepackage{xcolor}

\usepackage{jcappub}

\begin{document}

\title{The Atacama Cosmology Telescope: Constraints on Local Non-Gaussianity from the ACT Cluster Catalog}
\author[1]{Leonid Sarieddine,}
\author[2]{J. Richard Bond,}
\author[3,4]{Matt Hilton,}
\author[1,5]{Raul Jimenez,}
\author[6]{Arthur Kosowsky,}
\author[11]{Kavilan Moodley,}
\author[7,8,9]{Bernardita Ried Guachalla,}
\author[10]{Crist\'obal Sif\'on,}
\author[12]{Suzanne T. Staggs,}
\author[1,5]{Licia Verde,}
\author[13]{Edward J. Wollack,}

\affiliation[1]{ICCUB, University of Barcelona, Mart\' i i Franqu\` es, 1, E08028
Barcelona, Spain}

\affiliation[2]{Canadian Institute for Theoretical Astrophysics, University of Toronto, 60 St. George Street, Toronto, ON, M5S 3H8, Canada}

\affiliation[3]{Wits Centre for Astrophysics, School of Physics, University of the Witwatersrand, Private Bag 3, 2050, Johannesburg, South Africa}

\affiliation[4]{Astrophysics Research Centre, School of Mathematics, Statistics and Computer Science, University of KwaZulu-Natal, Durban 4001, South Africa}

\affiliation[5]{ICREA, Pg. Lluis Companys 23, Barcelona, 08010, Spain.} 

\affiliation[6]{Department of Physics and Astronomy, University of Pittsburgh, Pittsburgh, PA 15260, USA}

\affiliation[7]{Department of Physics, Stanford University, Stanford, CA, USA 94305-4085}

\affiliation[8]{Kavli Institute for Particle Astrophysics and Cosmology, 382 Via Pueblo Mall Stanford, CA 94305-4060, USA}

\affiliation[9]{SLAC National Accelerator Laboratory 2575 Sand Hill Road Menlo Park, California 94025, USA}

\affiliation[10]{Instituto de F\'isica, Pontif\'icia Universidad Catolica de Valparaiso, Casilla 4059, Valparaiso, Chile}

\affiliation[11]{Astrophysics Research Centre, School of Mathematics, Statistics and Computer Science, University of KwaZulu-Natal, Durban 4001, South Africa}

\affiliation[12]{Joseph Henry Laboratories of Physics, Jadwin Hall, Princeton University, Princeton, NJ, 08544, USA}

\affiliation[13]{NASA/Goddard Space Flight Center, Greenbelt, MD, USA 20771}

\emailAdd{leonid.sarieddine@icc.ub.edu; bond@cita.utoronto.ca; matt.hilton@wits.ac.za;   raul.jimenez@icc.ub.edu; 
kosowsky@pitt.edu;
moodleyk41@ukzn.ac.za;
bried@stanford.edu;
sifon@strw.leidenuniv.nl;
staggs@princeton.edu;
liciaverde@icc.ub.edu;
edward.j.wollack@nasa.gov;
}

\abstract{
We derive constraints on local-type primordial non-Gaussianity using the ACT DR6 Sunyaev--Zel'dovich cluster catalog. Modeling the redshift-- and mass--dependent number counts of 1,201 clusters in the 10,347 deg$^{2}$ Legacy region, and  accounting for  survey completeness, intrinsic SZ scatter, and a weak-lensing--calibrated mass bias, we compute theoretical abundances using the Log--Edgeworth halo mass function. Assuming $\Lambda$CDM with well-motivated external priors, we obtain $f_{\rm NL} = 55 \pm 125 \,(68\%~\mathrm{CL})$,
consistent with Gaussian initial conditions. These constraints probe comoving scales of $5$--$10\,\mathrm{Mpc}\,h^{-1}$, complementing CMB bispectrum and scale-dependent bias measurements, which do not reach such small scales. We also find evidence for a $16.4\%$ residual mass bias, which, although heavily informed by our adopted priors, plays a key role in matching observed and predicted counts but has negligible effect on $f_{\rm NL}$ constraints. We briefly discuss robustness of the results under relaxed priors and the prospects for next-generation SZ and lensing surveys to strengthen cluster-based tests of primordial non-Gaussianity.
}

\maketitle

\section{INTRODUCTION}

It has been long recognized that the cosmological abundance of rare objects is an excellent probe of the tail of the distribution of cosmological perturbations \cite{press1974formation, bardeen1986statistics,matarrese2000abundance}.  
In the standard paradigm of structure formation, halos correspond to peaks in the density fluctuations of the initial smoothed density field \cite{bardeen1986statistics, bond1991excursion}. Their statistical properties are characterized by the halo mass function which describes the number density of halos as a function of mass and redshift. In analytic approaches such as Press-Schechter \cite{press1974formation}, this abundance is modeled by linking halo formation to the probability of the linear smoothed density field exceeding a certain threshold; this makes the abundance extremely sensitive to any deviations from Gaussianity \cite{desjacques2010primordial,verde2010non, Dalal2008}. The rarest objects, i.e. those associated with the highest peaks of the density field, form the most massive dark matter halos, and at low redshift they are traced by the galaxy clusters who are therefore direct tracers of the high end of the halo population \cite{schmidt2007dark}. While the concept is simple, obtaining a precise and accurate quantification is not.

Challenges  reside on both the observational and theoretical front. On the observational side, it is necessary to obtain a faithful estimate of the mass of the target  dark matter halos \cite{pratt2019galaxy} (clusters of galaxies in this application), which is challenging. On the theoretical front, predicting the abundance of collapsed objects, which needs to be compared to observations, involves modeling highly non-linear physical processes \cite{zentner2007excursion}. While the latter could be done via numerical simulations, these are expensive, making   exploration of  the full parameter space  costly computationally and unmanageable in practice (e.g., \cite{borgani2011cosmological, yang2025ten}).

Primordial non-Gaussianity (PNG) — deviations from a purely Gaussian distribution of the initial curvature perturbations produced in the early Universe — is a powerful discriminator among inflationary scenarios. Among the first probes proposed in the literature was the abundance of rare, massive objects: galaxy clusters. Because the high-mass tail of the halo mass function samples extreme peaks of the initial density field, even small skewness or higher moments in the primordial distribution produce relatively large fractional changes in the number density of the most massive clusters, especially at high redshift \cite{matarrese2000abundance} (hereafter MVJ).

MVJ provided the first systematic analytic extension of the Press–Schechter framework to mildly non-Gaussian initial conditions. They computed the probability distribution for the smoothed density field including small non-Gaussian corrections and showed how these corrections map into a modified mass function for objects corresponding to very high peaks of the initial overdensity  distribution, and hence into altered abundances of high-$z$ objects. The key insight is that skewness (and other low-order moments) modifies the exponential sensitivity of the mass function’s tail, so cluster counts are exponentially sensitive to PNG parameters in the rare-object regime. 

A specific model for PNG, commonly parameterized by amplitudes such as $f_{\rm NL}$ for the bispectrum/skewness and $g_{\rm NL}$ for the trispectrum, gives corrections to the smoothed density probability distribution function and halo mass function $n(M,z)$ (e.g. \cite{loverde2008effects, loverde2011non, matarrese2000abundance}). Observed cluster counts (as a function of mass and redshift) can be compared to theoretical predictions that depend on cosmology, mass–observable relations, selection effects, and the PNG parameters \cite{allen2011cosmological, mak2012constraints}. Because the effect is strongest for the highest masses and redshifts, surveys that reliably identify massive clusters at $z\gtrsim1$ and that control selection biases and mass calibration systematics are the most valuable. 

Analytic formulae (including the MVJ prescription and later refinements \cite{loverde2008effects, loverde2011non, matarrese2000abundance}) require calibration against N-body simulations with non-Gaussian initial conditions \cite{Grossi2009, pillepich2010halo, wagner2012n}. Such comparisons validated the general approach and revealed necessary corrections, but also highlighted the impact of uncertainties in collapse thresholds, mass definition, and baryonic effects on mass estimates. Careful treatment of mass calibration, survey selection, and theoretical systematics is essential; otherwise apparent PNG signals can be mimicked by unaccounted systematics (e.g., \cite{mantz2015cosmology}).

Cluster abundance is highly complementary to CMB bispectrum and large-scale structure (scale-dependent bias) measurements: it probes different length scales and redshifts, and exploits different  signatures and physical processes. 

However, because of statistical and systematic limitations, constraints from cluster counts have typically been much weaker than the best CMB limits. The strongest, model-independent constraints on the local $f_{\rm NL}$ amplitude remain from the Planck CMB bispectrum: $f_{\rm NL}^{\rm local} = -0.9 \pm 5.1$ (68\% CL) \cite{akrami2020planck}. Galaxy-survey and multi-tracer large-scale structure  analyses are closing the gap and provide powerful scale-dependent tests \cite{slosar2008constraints,castorina2018primordial, Dalal2008}, but to date cluster-only constraints   have not been competitive. This situation is now changing with the latest ACT cluster catalog \cite{hilton2021atacama, aguena2025atacama}, which provides a substantially larger, higher-redshift, and more uniformly selected sample with improved mass calibration. 

Planck delivered the first large SZ-selected catalog of the most massive galaxy clusters \cite{ade2016planck}. the higher angular resolution and lower noise maps of Atacama Cosmology Telescope \cite{aguena2025atacama} and South Pole telescope \cite{bleem2015galaxy} have since increased SZ cluster catalog sizes substantially for a given sky area. In particular, recent ACT releases \cite{aguena2025atacama, hilton2021atacama} constitute the largest SZ-selected galaxy cluster catalog to date. It is this cluster catalog that is going to be the focus of this paper. 

Galaxy cluster catalogs based on blind detection of the thermal Sunyaev-Zel'dovich (SZ) effect yield the largest nearly-mass-limited cluster samples because their selection is relatively simple and highly complete above a fixed mass threshold \cite{mroczkowski2019astrophysics}. The SZ signal is almost independent of redshift for a given cluster mass \cite{birkinshaw1999sunyaev}, enabling detection of essentially all massive clusters throughout the Hubble volume. In addition, the integrated SZ distortion provides a relatively direct and robust proxy for total cluster mass \cite{kravtsov2006new}.

This paper is organized as follows. In section~\ref{sec:2} we begin with a description of the ACT cluster catalog. We then review the literature on the halo mass function and specify our PNG model in section~\ref{sec:3}. In section~\ref{sec:4} we describe the pipeline for computing the predicted number counts, including the treatment of mass bias, the construction of the mass–scattering kernel, and related systematics. With these ingredients in place,  we specify our priors and the likelihood function in section~\ref{sec:5} and then present the results  and constraints on $f_{\rm NL}$ in section \ref{sec:results}. Finally, the concluding section \ref{sec:conclusions} discusses how our results compare to the literature. Appendix~\ref{sec:appendix} reviews analytic mass functions from the literature. For completeness, we also evaluate their predicted cluster number counts for the ACT data set, although these mass functions are not expected to provide an optimal description for the reasons discussed in the appendix.

\section{ACT DR6 SZ CLUSTER CATALOG: DATA, SELECTION, AND MASS PROXIES}
\label{sec:2}
The ACT DR6 Sunyaev--Zel'dovich (SZ) cluster catalog \cite{hilton2021atacama,aguena2025atacama}, is a large, homogeneously-selected sample of galaxy clusters identified through their thermal SZ (tSZ) signatures in multi-frequency maps from the Atacama Cosmology Telescope (ACT). The SZ effect traces the integrated electron pressure along the line of sight and, to a good approximation, its surface brightness is redshift independent. As a result, an SZ survey is approximately mass-limited over a wide redshift range and efficiently selects the most massive halos at all epochs \citep{carlstrom2002cosmology}.

The DR6 release combines roughly a decade and a half of ACT observations into coadded maps in three frequency bands centered at 98, 150, and 220\,GHz over a sky area of $16{,}293~\mathrm{deg}^2$ (with a high-purity “Legacy” region of $10{,}347~\mathrm{deg}^2$). Here, we briefly summarize  key information from \cite{hilton2021atacama,aguena2025atacama} that is relevant for this work. Cluster candidates are extracted with multi-frequency matched spatial filters built from a universal pressure-profile (UPP) template \cite{arnaud2010universal} which encodes the expected SZ spatial morphology and the thermal SZ spectral dependence. The catalogue lists all detections above a fixed signal-to-noise threshold (SNR\,$\ge4$ at a filter scale of $2.4'$ by default; many cosmology analyses adopt SNR\,$>5.5$ and restrict to the Legacy footprint for uniform completeness), and then confirms them using optical/IR imaging (primarily from the DES, HSC, DESI Legacy surveys) \cite{rykoff2016redmapper, oguri2018optically, rykoff2014redmapper}. Spectroscopic redshifts are included when available, and otherwise photometric redshifts with quoted uncertainties are assigned. the sample spans the redshift range $0< z < 2.0$ with a median around $z = 0.58$, including 1,166 clusters at $z>1$  and 121 at $z>1.5$.

Each catalog entry includes celestial coordinates, the matched-filter SNR, the filter scale parameter, the central or filtered Compton-$y$ amplitude with measurement uncertainty, and redshift information and quality flags. Where possible, the catalog also provides footprint flags indicating whether a cluster lies inside the coverage of specific imaging or weak-lensing datasets. External optical cluster catalogs are cross-matched to supply redshifts, but richness estimates (e.g.,\ $\lambda$ from redMaPPer \cite{rykoff2016redmapper, oguri2018optically, rykoff2014redmapper}) are not reported as part of the ACT DR6 release.

The ACT DR6 catalog reports several different mass estimates tied to the spherical overdensity mass within $R_{500c}$, the radius enclosing $500$ times the critical density. The quantity \texttt{M500cUncorr} is a mass derived from the SZ amplitude via the UPP scaling relation without applying a Bayesian Eddington correction; its quoted errors reflect measurement noise (on the SZ amplitude and the redshift) propagated through the scaling relation. The quantity \texttt{M500c} applies a Bayesian treatment that corrects for Eddington bias---the combination of the steeply falling halo mass function and the intrinsic log-normal scatter of the observable--mass relation. In DR6, both of these use an updated SZ--mass scaling relation normalized to $10^{A_0}=3.0\times10^{-5}$ (down from $4.95\times10^{-5}$ in DR5), which lowers the overall mass scale by $60\%$ compared to DR5 and brings it into agreement with weak-lensing calibrations. In addition, DR6 reports masses on the older A10 normalization (\texttt{A10\_M500c}, with corresponding uncorrected versions) to facilitate comparison with Planck and previous ACT releases. Unlike DR5, no separate \texttt{M500cCal} column is needed: the default \texttt{M500c} values are already on a weak-lensing calibrated mass scale, consistent with results from  KiDS and HSC \citep{robertson2024act, miyatake2019weak, PhysRevD.110.103006}, implying an average mass bias of $1-b_{\rm fid} \footnote{Note that $b_{\rm fid}$ is just the standard mass bias. We will later introduce a residual mass bias which we will denote by $b$.} = 0.65$. 
Hereafter we adopt this quantity as  the cluster total mass  and refer to it as $M_{\rm500c}$.

In addition, the SZ--mass relation is assumed to exhibit an intrinsic log-scatter of $\sigma_{\ln M}\simeq0.185$. Cosmological inference analyses typically treat the weak-lensing mass calibration uncertainty as a nuisance parameter, while $\sigma_{\ln M}$ controls the parameter distribution width rather than the mean.

Survey completeness is characterized using a semi-analytic approach, validated through end-to-end simulations of the cluster finder. The DR6 catalog is approximately $90\%$ complete for SNR $\ge5.5$ detections above $M_{500c}\simeq 5\times 10^{14}\,M_\odot$ at $z\simeq0.2$ (within the Legacy footprint); completeness falls at lower mass and depends on redshift and the depth of the local map. The release provides the ingredients to build a selection function $C(M,z)$ which quantifies the detection probability as a function of mass and redshift; this makes it possible to account for selection effects when comparing theoretical predictions of cluster abundance to the observed data. Because the ACT selection is driven primarily by mass (through the SZ signal) and only weakly by redshift, the resulting sample is close to mass-limited across the survey volume, which is advantageous for cosmological inference. In this work, we restrict our analysis to the ACT DR6 Legacy footprint.

\section{HALO MASS FUNCTION}
\label{sec:3}

For our purposes, the halo mass function quantifies the comoving number density of clusters as a function of mass and redshift and is a central ingredient for connecting large-scale-structure data to cosmological parameters \cite{press1974formation,bond1991excursion,sheth1999large}. In the excursion-set picture, halos form when the smoothed linear overdensity $\delta_R(\mathbf{x},z)$ first exceeds a collapse threshold $\delta_c$ as the smoothing scale $R$ is decreased \cite{bond1991excursion}. Each smoothing scale $R$ corresponds to a characteristic mass
$M = \frac{4\pi}{3} R^3 \bar{\rho}_m \,,$ where $\bar{\rho}_m = \Omega_{m,0}\,\rho_{\rm crit,0}$ is the mean matter density with $\rho_{\rm crit,0} = \frac{3H_0^2}{8\pi G}$, so reducing the smoothing scale probes the abundance of increasingly smaller objects. Writing the mass function in terms of the variance $\sigma^2(M,z)$ of the linear density field smoothed with a (real-space) top-hat filter, a convenient and widely used form, gives \cite{press1974formation,sheth1999large}
\begin{equation}
\frac{\mathrm{d}n}{\mathrm{d}\ln M}(M,z) \;=\; \frac{\bar\rho_m}{M}\,f(\sigma)\,\left|\frac{\mathrm{d}\ln\sigma^{-1}}{\mathrm{d}\ln M}\right|
\;=\; \frac{\bar\rho_m}{M}\,\psi(\nu)\,,
\end{equation}
where $\nu\equiv \delta_c/\sigma(M,z)$ is the peak height, $f(\sigma)$ is the multiplicity function expressed in terms of the rms fluctuation $\sigma$ (dropping the explicit dependence on $M$ and $z$), and $\psi(\nu)\equiv f(\sigma)$ after the change of variables $\nu=\delta_c/\sigma$. The spherical-collapse threshold is $\delta_c\simeq 1.686$ in an Einstein--de Sitter cosmology (with mild cosmology and redshift dependence in $\Lambda$CDM). The variance is
\begin{equation}
\sigma^2(M,z) \;=\; \frac{1}{2\pi^2}\int_0^\infty k^2 P_\mathrm{L}(k,z)\,W^2(kR)\,\mathrm{d}k\,,
\end{equation}
with $P_\mathrm{L}(k,z)$ the linear power spectrum, $W(kR)$ the Fourier transform of the top-hat filter $W(x)=3(\sin x-x\cos x)/x^3$, and $R=(3M/4\pi\bar\rho_m)^{1/3}$ \cite{Peebles1980,LiddleLyth2000}.

For spherical-overdensity halo definitions, Tinker et al. \cite{Tinker2008} provided a widely used halo mass function fit to N-body simulations that we adopt here:
\begin{equation}
f_\mathrm{Tinker08}(\sigma) \;=\; A\left[\left(\frac{\sigma}{b}\right)^{-a}+1\right]\exp\!\left(-\frac{c}{\sigma^2}\right),
\label{Tinker08}
\end{equation}
with parameters $A, a, b, c$ tabulated as functions of redshift and overdensity (e.g., $\Delta=200$--3200 relative to the mean). While \cite{Watson2013} extended calibrations to higher redshift, and  subsequent works explored cosmology and baryonic effects (e.g. \cite{Bocquet2016,cui2014effect,castro2021impact}), these works show that
baryonic effects mostly lead to a modification of individual halo masses without significantly altering the shape of the mass function; in our work we use the ACT calibrated masses which already include these effects. 

Deviations from Gaussianity alter the abundance of rare massive halos and hence the halo mass function. MVJ \cite{matarrese2000abundance} extended the Press–Schechter formalism to the non-Gaussian case and provided a first order analytic expression for the non-Gaussian halo mass function using the saddle point approximation, designed to provide a good description in the limit of $\nu\gg 1$ i.e., very  high peaks or very rare objects. An alternative approach was used by LoVerde et al.\ (LV, \cite{loverde2008effects}) using an Edgeworth expansion, which is designed to apply to less extreme $\nu$.
All analytic prescriptions agree qualitatively with simulations but require calibration. 

For example, \cite{Grossi2009, wagner2012n} showed that excellent agreement with $N$-body results is obtained by rescaling the collapse threshold, 
$\delta_c\rightarrow \delta_c\sqrt{q}$ with $q\simeq 0.75$.
In particular, they proposed that the full non-Gaussian halo mass function should be written as:
 \begin{equation}
n_{\mathrm{NG}}(M,z,f_{\mathrm{NL}}) \;=\;
n^{\mathrm{sim}}_{\rm G}(M,z) \;\times\; R_{\mathrm{NG}}(M,z)
\label{non-Gaussian hmf}
\end{equation}
where $n^{\mathrm{sim}}_{G}(M,z)$ is the Gaussian halo mass function calibrated to N-body simulations (here we adopt \cite{Tinker2008}), and $R_{\mathrm{NG}}(M,z)$ is the non-Gaussian multiplicative factor defined as the ratio of the non-Gaussian to Gaussian theory mass function:
\begin{equation}
R_{\mathrm{NG}}(M,z)\equiv \frac{n^{\rm th}_{\rm NG}(M,z,f_{\rm NL})}{n_{\rm G}^{\rm th}(M,z)}\,.
\end{equation}

For high enough $|f_{\rm NL}|$ values, both the plain Edgeworth expansion of LV and the MVJ approaches can become negative (or imaginary in MVJ’s case) in the rare–event tail, producing
unphysical probabilities for very massive haloes. For more details, see Appendix \ref{sec:appendix}.

LoVerde \& Smith \cite{loverde2011non}, addressed these limitations by developing the Log-Edgeworth halo mass function.  Instead of applying the Edgeworth
series directly to the probability distribution, they expanded the
logarithm of the correction factor, which has the practical
advantage of enforcing positivity of the final result.  This log-resummed approach improves convergence in the high-mass tail, where the naive Edgeworth form can fail and systematically incorporates not only skewness (controlled by $f_{\rm NL}$) but also
kurtosis and variance corrections arising from higher-order parameters such as $g_{\rm NL}$ and $\tau_{\rm NL}$.  In doing so, it provides a broader and more accurate analytic description of halo abundances across different models of primordial non-Gaussianity.

To be more specific, they describe the non-Gaussian statistics of the smoothed density field using reduced cumulants \(\kappa_n(M)\). For the local family of initial conditions, useful fits for the skewness and kurtosis are
\begin{gather}
\kappa_3(M) \;\approx\;
f_{\rm NL}\,(6.6\times 10^{-4})
\Big[\,1 - 0.016\,\ln\!\big(M/(h^{-1}M_\odot)\big)\Big],
\label{eq:kappa3_fit}
\\[6pt]
\begin{gathered}
\kappa_4(M) \;\approx\;
g_{\rm NL}\,(1.6\times 10^{-7})
\Big[\,1 - 0.021\,\ln\!\big(M/(h^{-1}M_\odot)\big)\Big]
\\[2pt]
\quad
+ \frac{\tau_{\rm NL}}{(6/5)^2}
\Big[\,(6.9\times 10^{-7})
\Big(1 - 0.021\,\ln\!\big(M/(h^{-1}M_\odot)\big)\Big)
+ 48\,\Delta_\Phi^2 \ln(L/L_0)\Big],
\end{gathered}
\label{eq:kappa4_fit}
\end{gather}
and we write the total variance as a Gaussian part times a correction,
\begin{gather}
\sigma^2(M) = \sigma_G^2(M)\,[\,1+\kappa_2(M)\,], \label{eq:sigma_split}\\[2pt]
\kappa_2(M) \approx
\frac{\tau_{\rm NL}^2}{(6/5)^4 f_{\rm NL}^2}
\Big[(4.0\times 10^{-8})\big(1 - 0.021\,\ln(M/(h^{-1}M_\odot))\big)
+ 4\,\Delta_\Phi^2 \ln(L/L_0)\Big]. \label{eq:kappa2_fit}
\end{gather}
Here \(L_0=1600\,h^{-1}\mathrm{Mpc}\) and the amplitude of the primordial fluctuations is given by \(\Delta_\Phi^2=\tfrac{9}{25}\Delta_\zeta^2\simeq 8.72\times 10^{-10}\) for the fiducial cosmology. Here \(\Delta_\Phi^2\) and \(\Delta_\zeta^2\) refer to the potential and curvature perturbation fluctuations respectively.

We define the collapse threshold\footnote{ In an Einsten de Sitter Universe  the collapse threshold is 1.68, in $\Lambda$CDM this acquires a weak  redshift dependence which we ignore as its effect is small. Importantly, the value for the collapse threshold we adopt already includes the correction factor calibrated on simulations as done in \cite{loverde2011non}.} \(\delta_c\simeq 1.42\) and \(\nu_c(M)=\delta_c/\sigma(M)\).
The cumulative collapsed fraction is expanded as
\begin{equation}
F(M)=F_0(M)+F_1(M)+F_2(M),
\end{equation}
with
\begin{align}
F_0(M) \;=\;& \tfrac12\,\mathrm{erfc}\!\Big(\nu_c(M)/\sqrt{2}\Big),
\label{eq:F0}
\\[4pt]
F_1(M) \;=\;& \frac{1}{\sqrt{2\pi}}\,\exp\!\Big[-\tfrac12\,\nu_c^2(M)\Big]\,
\Bigg[\frac{\kappa_3(M)}{6}\,H_2\!\big(\nu_c(M)\big)\Bigg],
\label{eq:F1}
\\[4pt]
F_2(M) \;=\;& \frac{1}{\sqrt{2\pi}}\,\exp\!\Big[-\tfrac12\,\nu_c^2(M)\Big]\,
\Bigg[\frac{\kappa_2(M)}{2}\,H_1\!\big(\nu_c(M)\big)
\nonumber\\
&\qquad\qquad
+ \frac{\kappa_4(M)}{24}\,H_3\!\big(\nu_c(M)\big)
+ \frac{\kappa_3^2(M)}{72}\,H_5\!\big(\nu_c(M)\big)\Bigg],
\label{eq:F2}
\end{align}
where the Hermite polynomials satisfy \(H_n(\nu)=(-1)^n e^{\nu^2/2}\,\tfrac{\mathrm{d}^n}{\mathrm{d}\nu^n}e^{-\nu^2/2}\), i.e.
\begin{align}
&H_1=\nu,\quad
H_2=\nu^2-1,\quad
H_3=\nu^3-3\nu,\quad
H_4=\nu^4-6\nu^2+3,\nonumber\\
&H_5=\nu^5-10\nu^3+15\nu,\quad
H_6=\nu^6-15\nu^4+45\nu^2-15.
\label{eq:Hermites}
\end{align}
Taking derivatives with respect to \(M\),
\begin{equation}
F_0'(M) \;=\; -\,\nu_c'(M)\,\frac{1}{\sqrt{2\pi}}\,\exp\!\Big[-\tfrac12\,\nu_c^2(M)\Big],
\label{eq:F0prime}
\end{equation}
\begin{equation}
F_1'(M) \;=\;
F_0'(M)\left[\frac{\kappa_3(M)}{6}\,H_3\!\big(\nu_c(M)\big)
- \frac{\kappa_3'(M)}{6\,\nu_c'(M)}\,H_2\!\big(\nu_c(M)\big)\right],
\label{eq:F1prime}
\end{equation}
\begin{align}
F_2'(M) \;=\; F_0'(M)\,\Bigg[&
\frac{\kappa_2(M)}{2}\,H_2\!\big(\nu_c(M)\big)
+ \frac{\kappa_4(M)}{24}\,H_4\!\big(\nu_c(M)\big)
\nonumber\\
&+ \frac{\kappa_3^2(M)}{72}\,H_6\!\big(\nu_c(M)\big)
- \frac{\kappa_2'(M)}{2\,\nu_c'(M)}\,H_1\!\big(\nu_c(M)\big)
\nonumber\\
&
- \frac{\kappa_4'(M)}{24\,\nu_c'(M)}\,H_3\!\big(\nu_c(M)\big)
- \frac{\kappa_3(M)\,\kappa_3'(M)}{36\,\nu_c'(M)}\,H_5\!\big(\nu_c(M)\big)\Bigg].
\label{eq:F2prime}
\end{align}
The Log-Edgeworth prescription keeps terms to the same order in \(\ln F\),
\begin{equation}
\ln F(M)\;\approx\;\ln F_0(M)\;+\;\frac{F_1(M)}{F_0(M)}\;+\;\frac{F_2(M)}{F_0(M)}
\;-\;\frac12\left(\frac{F_1(M)}{F_0(M)}\right)^{\!2}.
\label{eq:lnFexp}
\end{equation}
From this, the ratio of the non-Gaussian to Gaussian mass functions is
\begin{align}
R_{\rm NG}(M,z,\kappa_2,\kappa_3,\kappa_4)  \;\approx\;&
\exp\!\Bigg[
\frac{F_1}{F_0} + \frac{F_2}{F_0} - \frac12\!\left(\frac{F_1}{F_0}\right)^{\!2}
\Bigg]\,\times
\nonumber\\
&
\Bigg[
1 + \frac{F_1'}{F_0'} + \frac{F_2'}{F_0'}
- \left(\frac{F_1}{F_0}\right)\!\left(\frac{F_1'}{F_0'}\right)
- \left(\frac{F_1}{F_0} + \frac{F_2}{F_0}\right)
+ \left(\frac{F_1}{F_0}\right)^{\!2}
\Bigg],
\label{eq:logEdge_ratio}
\end{align}
with \(F_0,F_1,F_2\) and their derivatives given in Eq.~\eqref{eq:F0}--\eqref{eq:F2prime}. Note that Eq.~\eqref{eq:kappa3_fit}--\eqref{eq:logEdge_ratio} all come from Ref.~\cite{loverde2011non}.

In our analysis we restrict to the local-type bispectrum with $g_{\rm NL}=0$, and we assume the single-source relation for the trispectrum,
\begin{equation}
\tau_{\rm NL} \;=\; \left( \frac{6 f_{\rm NL}}{5} \right)^{2},
\end{equation}
which saturates the Suyama-Yamaguchi inequality \cite{suyama2008non} and holds in models where the primordial curvature perturbation is generated by a single degree of freedom. Therefore in this application the dependence of  $R_{\rm NG}$ is only on $z$, halo mass and $f_{\rm NL}$.

To summarize, we first calculate the $n^{\mathrm{sim}}_{\rm G}(M,z)$ halo mass function using Eq.~(\ref{Tinker08}) with a critical spherical overdensity of 500, then we use Eq.~(\ref{eq:logEdge_ratio}) as the $R$ factor for the non-Gaussian to Gaussian ratio.

\section{NUMBER COUNTS}
\label{sec:4}
\subsection{Observed number counts from the catalog}

The catalog provides the mass estimate $M_{\rm 500c}$ (converted to $M_\odot$ by multiplying by $10^{14}$),
the redshift $z$, and signal to noise ratio (SNR). We apply the cut $\mathrm{SNR} > 5.5$ as well as the legacy footprint\footnote{In practice we only consider clusters that satisfy the mask  
\[
\textbf{Criteria} = (\mathrm{SNR} > 5.5) \;\wedge\; (\mathrm{flags} = 0) \;\wedge\; (\mathrm{footprint\_legacy} = 1).
\]}, leaving  3758 clusters  \cite{aguena2025atacama}. We  bin them into
redshift bins of width $0.1$ from $z=0.2$ to $z=2.1$ and  bin the logarithmic observed-masses in bins of width 0.1 (in units of $\log_{10}$), from $\log_{10}M = 14.6$ to $\log_{10}M = 15.7$. The adopted  mass and redshift lower limits ensure that the data has a completeness of over $90\%$, leaving $1201$ clusters. 
This high completeness threshold ensures that the completeness correction to the resulting PNG constraints remains small.

\subsection{Mass systematics}

ACT infers masses from the SZ signal via the UPP \cite{arnaud2010universal} and the DR6 mass–observable relation \cite{aguena2025atacama}
\begin{equation}
\tilde{y}_0 \;=\; 10^{A_0}\,E(z)^2
\left(\frac{M_{500\mathrm{c}}}{M_{\rm pivot}}\right)^{1+B_0}\,
Q(\theta_{500\mathrm{c}})\,f_{\rm rel}(M_{500\mathrm{c}},z),
\label{eq:DR6-scaling}
\end{equation}
Here $\tilde{y}_0$ is the matched–filter central Compton-$y$ amplitude, extracted from a map filtered at 2.4 arcminute filter scale. $M_{500\mathrm{c}}$ is the mass enclosed within a sphere of overdensity 500 times the critical density, $\theta_{500\mathrm{c}}$ is the corresponding angular radius, $A_0$ and $B_0$ are the normalization and slope parameters of the scaling relation with $B_0=0.08$, $M_{\rm pivot}$ is the chosen pivot mass with $M_{\rm pivot}=3\times10^{14}\,M_\odot$. The redshift dependence is $E(z)=[\Omega_m(1+z)^3+\Omega_\Lambda]^{1/2}$ with $\Omega_m$ and $\Omega_\Lambda$ being the present-day matter and vacuum energy density fractions. $Q(\theta_{500\mathrm{c}})$ is the filter–mismatch correction, also provided in the catalog, and $f_{\rm rel}(M_{500\mathrm{c}},z)$ accounts for relativistic corrections to the thermal SZ spectrum (as in \cite{itoh1998relativistic}).  The normalization of DR6 is
\[
10^{A_0}=3.0\times10^{-5},
\]
replacing the older A10 value $4.95\times10^{-5}$ used in prior ACT/Planck catalogs \cite{arnaud2010universal, hilton2021atacama}. This change (new normalization $\simeq60\%$ of A10) lifts the mass scale at fixed $\tilde{y}_0$ so that ACT SZ masses align with stacked WL measurements of ACT clusters.

The probability distribution for the cluster mass $M_{500\mathrm{c}}$ given the cluster redshift and the central Compton-$y$ amplitude is denoted by $P(M_{500\mathrm{c}}\!\mid\!\tilde{y}_0,z)$. 
To correct for the Eddington bias, ACT DR6 adopts the posterior \cite{aguena2025atacama,hilton2021atacama}
\begin{equation}
P(M_{500\mathrm{c}}\!\mid\!\tilde{y}_0,z)\;\propto\;
P(\tilde{y}_0\!\mid\!M_{500\mathrm{c}},z)\,P(M_{500\mathrm{c}}\!\mid\!z),
\label{eq:DR6-posterior}
\end{equation}
to infer the mass $M_{500\mathrm{c}}$
with  the conditional probability $P(\tilde{y}_0\!\mid\!M_{500\mathrm{c}},z)$ as a log–normal likelihood around a mean given by Eq.~\ref{eq:DR6-scaling} and assume an intrinsic scatter in $\tilde{y}_0$ of $\sigma_{\rm int}=0.2$ (based on \cite{hasselfield2013atacama}) which is the cluster-to-cluster physical dispersion, or the width of the error-of-the-mean relation for the proxy at fixed true mass. It contributes to the spread in observed values, and thus to the likelihood of drawing a biased mass. For $P(M_{500\mathrm{c}}\!\mid\!z)$  they adopt a Tinker halo–mass–function \cite{Tinker2008} prior evaluated with CCL (v3.1.2) under $\sigma_8=0.80$. Reported mass uncertainties include measurement noise and intrinsic scatter; they do not propagate uncertainties on the scaling‐relation parameters. 

To compute predicted counts, we need the mass response function $P(M_{\rm 500c}^{\rm obs}\mid M_{\rm 500c}^{\rm true})$ which is the probability distribution of observing a cluster with mass $M_{\rm 500c}^{\rm obs}$ given its true mass $M_{\rm 500c}^{\rm true}$. To be clear, we take $M_{\rm 500c}^{\rm obs}$ to be the mass inferred by ACT via Eq.~\ref{eq:DR6-posterior} which depends on the observed SZ signal; meanwhile $M_{\rm 500c}^{\rm true}$ refers to the true mass of the cluster and is the mass on which the predicted cluster mass function depends, i.e. $n_{\rm NG}(M_{\rm 500c}^{\rm true})$. $P(M_{\rm 500c}^{\rm obs}\mid M_{\rm 500c}^{\rm true})$ is the response function that maps between the two masses and is essential for calculating the predicted number counts. To that end, we first derive the log-normal intrinsic scatter in $M_{\rm 500c}$. We note that the mean $\tilde y_0-M_{\rm 500c}$ relation is locally linear in logs with slope
\[
s \;\equiv\; \frac{d\ln \tilde y_0}{d\ln M_{\rm 500c}}.
\]
For a log-normal relation, the corresponding scatter in mass is
\[
\sigma_{\ln M_{\rm 500c}} \;\simeq\; \frac{\sigma_{\ln \tilde y_0}}{\,s\,}.
\]

 From Eq.~\ref{eq:DR6-scaling} we see that the log-slope is $s= 1+B_0$ with $B_0=0.08$, hence $s=1.08$ and thus 
\begin{align}
    \sigma_{\ln M_{\rm 500c}}= \frac{0.2}{1.08} \approx 0.185
\label{eq: intrinsic scatter}
\end{align}

It is well known that SZ masses are in general biased low when compared to the weak lensing masses (see e.g. \cite{ade2016planck}). In particular \cite{robertson2024act},
\begin{align}
\frac{\langle M_{500\mathrm{c}}^{\rm A10}\rangle}{\langle M_{500\mathrm{c}}^{\rm WL}\rangle} = 1-b_{\rm fid} = 0.65 \pm 0.05.
\label{eq:ACTbias}
\end{align}
The catalog mass referred to as $M_{\rm 500c}$ already incorporates this mass bias; since we are using the calibrated $M_{500\rm c}$ masses we do not expect a mass bias, but we allow the model to detect any potential residual bias by  assuming a prior on the residual mass bias centered around zero with a width $\frac{0.05}{0.65} \approx 0.077$ (see Eq.~\ref{eq:ACTbias}): %as follows:
\begin{equation}
\ln(1-b) \sim\mathcal{N}(0,0.077^2).
\label{eq:massbias}
\end{equation}

Putting everything together, we assume that the observed mass at fixed true mass is log-normally distributed and has the following specific form:
\begin{equation}
\ln M_{\rm 500c}^{\rm obs} \sim\mathcal{N}\big( \ln M_{\rm 500c}^{\rm true} + \ln (1-b) ,\sigma_{\ln M_{\rm 500c}} \big).
\label{eq:lnmap}
\end{equation}

Note that we assume a constant residual mass bias (see Eq.~\ref{eq:massbias}). Ref.~\cite{robertson2024act} Explicitly states that it finds no evidence of strong mass dependency of the mass bias. On the other hand, recent ACT weak-lensing calibrations have improved precision, and a DES year-3 calibration \cite{shin2025weak} finds significant redshift dependence. So while recent analyses do suggest the mass bias evolves with redshift, this doesn't necessarily imply a redshift dependence of the residual correction. We therefore adpot a constant residual bias as an effective description.

To calculate the total bias between the uncalibrated mass and the true mass, note that $\langle M_{500\mathrm{c}}^{\rm A10}\rangle = \langle M^{\rm obs}_{\rm 500c} \rangle$ hence using Eq.~\ref{eq:ACTbias} and the mean of Eq.~\ref{eq:lnmap} we obtain:

$$\langle M_{500\mathrm{c}}^{\rm A10}\rangle = (1-b_{\rm fid})(1-b)\langle M^{\rm true}_{\rm 500c} \rangle$$.

\subsection{Mass response function $P(M_{\rm 500c}^{\rm obs}\mid M_{\rm 500c}^{\rm true})$}

This subsection specifies the forward model that maps a set of ``true'' halo masses to binned observed masses via a log-normal response calibrated on the Atacama Cosmology Telescope (ACT) cluster catalog.  From Eq.~\ref{eq:lnmap} we can now explicitly write down the probability distribution of observing a cluster with mass $M_{\rm 500c}^{\rm obs}$ given its true mass is $M_{\rm 500c}^{\rm true}$:
\begin{equation}
P(M_{\rm 500c}^{\rm obs}|M_{\rm 500c}^{\rm true}) = \frac{1}{\sqrt{2 \pi \sigma_{\ln M_{\rm 500c}}^2}}\exp \bigg[ -\frac{\big(\ln M_{\rm 500c}^{\rm obs} - \ln M_{\rm 500c}^{\rm true}- \ln(1-b)\big)^2}{2\sigma_{\ln M_{\rm 500c}}^2}\bigg]\equiv \frac{e^{-{\cal Y}^2(M_{\rm 500c}^{\rm obs})}}{\sqrt{2 \pi \sigma_{\ln M_{\rm 500c}}^2}}
\end{equation}

Note that the probability of a cluster of true mass $M_{\rm 500c}^{\rm true}$ being observed in the  observed mass bin $\Delta M_m=[M_m^L, M_m^U]$ is

\begin{equation}
P\!\left(M_{\rm 500c}^{\rm true}  \in
\Delta M_m\right)
=
\int_{\ln M_m^L}^{\ln M_m^U}
P(M_{\rm 500c}^{\rm obs}|M_{\rm 500c}^{\rm true}) d\ln M_{\rm 500c}^{\rm obs}=\frac{1}{2}\left[\erf({\cal Y}(M_m^U))-\erf({\cal Y}(M_m^L))\right]
\end{equation}
where $\erf$ is the error function.

\subsection{Predicted counts}

Let $\mu_{r,m}$ denote
the predicted ensemble-averaged number of objects in the redshift bin $r$, denoted by $\Delta z_r$, and the mass bin $m$, denoted by $\Delta M_m$.  The predicted counts are given by (see also, e.g., \cite{mana2013combining, mak2012constraints}),

\begin{align}
 \mu_{r,m}
& = 
\Delta\Omega
\int_{\rm \Delta z_r} dz\;
\frac{d^2V(z)}{dz\,d\Omega}
\int_{0}^{\infty} d\ln M_{\rm 500c}^{\rm true}\;
n_{\rm NG}(M_{\rm 500c}^{\rm true},z)
 \\ \nonumber
&  \times \int_{\rm \Delta M_m } d\ln M_{\rm 500c}^{\rm obs}\;
P(M_{\rm 500c}^{\rm obs}|M_{\rm 500c}^{\rm true}) C(M_{\rm 500c}^{\rm true},z)  \; 
\label{eq: number counts1}
\end{align}

\begin{equation}
\mu_{r,m}
=
\Delta\Omega
\int_{\rm \Delta z_r} dz\;
\frac{d^2V(z)}{dz\,d\Omega}
\int_{0}^{\infty} d\ln M_{\rm 500c}^{\rm true}\;
n_{\rm NG}(M_{\rm 500c}^{\rm true},z) C(M_{\rm 500c}^{\rm true},z)
P\!\left(M_{\rm 500c}^{\rm true}  \in
\Delta M_m\right)
\label{eq: number counts 2}
\end{equation}

\noindent where $n_{NG}(M_{\rm 500c}^{\rm true},z)$ is given explicitly by Eq.~\ref{non-Gaussian hmf} and \ref{eq:logEdge_ratio} and $C(M_{\rm 500c}^{\rm true},z)$ is the completeness function. 

As usual, assuming flatness\footnote{We directly use the astropy library with Planck18 cosmology to compute this.},
$$\frac{d^{2}V(z)}{dz \, d\Omega} \;=\; \frac{c}{H(z)}\int_{0}^{z} \frac{c}{H(z')} \, dz'$$ and $\Delta\Omega$ corresponds to the  survey area of 10347 deg$^2$. Note that the actual number of objects in each bin differs due to cosmic variance and Poisson noise. For rare objects, Poisson noise will dominate.

\section{PRIORS AND LIKELIHOODS}
\label{sec:5}
The free parameters in the analysis are $\theta=(f_{\rm NL}, \ln(1-b), \sigma_8, h, \Omega_m)$. The main parameter of interest is $f_{\rm NL}$, while the others are treated as nuisance parameters which we marginalize over to give the $f_{\rm NL}$ posterior. We use a uniform prior for $f_{\rm NL}$ and Gaussian priors for the rest, with stated means and standard deviations (see baseline case in table \ref{tab:priors}).
The justification for the  baseline priors is as follows: the mass bias prior comes from \cite{robertson2024act} and  Eq.~\ref{eq:ACTbias} with the intrinsic scatter from Eq.~\ref{eq: intrinsic scatter}; the cosmological parameters $\sigma_8$, $h$, $\Omega_m$ are assumed to have priors given by the Planck 2018 posterior constraints \cite{aghanim2020planck}.

In the following sections, we will also show results for two other cases: partially relaxed priors on the cosmological parameters by doubling the prior width, and fully relaxed priors where in addition we double the prior width in the residual mass bias and the intrinsic scatter $\sigma_{\ln M}$. Table \ref{tab:priors} summarizes all three cases.

\begin{table}[htbp]
\centering
\caption{Summary of prior choices for the three analysis setups. $\mathcal{U}$ denotes uniform prior and $\mathcal{N}$ denotes Gaussian prior with stated mean and variance. The baseline priors are from Planck 2018 Ref.~\cite{aghanim2020planck} and Eqs.~\ref{eq:ACTbias}, \ref{eq: intrinsic scatter}. In the partially relaxed prior we double the width of the cosmological parameters priors, and in the fully relaxed priors we also double the width of the residual mass bias and mass scatter priors.}
\begin{tabular}{lccc}
\hline\hline
Parameter & Baseline & Partially--relaxed & Fully--relaxed \\[2pt]
\hline
$f_{\rm NL}$ 
& $\mathcal{U}(-1000,\,1000)$ 
& $\mathcal{U}(-1000,\,1000)$ 
& $\mathcal{U}(-1000,\,1000)$ \\[6pt]

$\ln(1-b)$ 
& $\mathcal{N}(0,\,0.077^{2})$
& $\mathcal{N}(0,\,0.077^{2})$
& $\mathcal{N}(0,\,0.154^{2})$ \\[6pt]

$h$
& $\mathcal{N}(0.674,\,0.005^{2})$
& $\mathcal{N}(0.674,\,0.01^{2})$
& $\mathcal{N}(0.674,\,0.01^{2})$ \\[6pt]

$\sigma_{8}$
& $\mathcal{N}(0.811,\,0.006^{2})$
& $\mathcal{N}(0.811,\,0.012^{2})$
& $\mathcal{N}(0.811,\,0.012^{2})$ \\[6pt]

$\Omega_{m}$
& $\mathcal{N}(0.315,\,0.007^{2})$
& $\mathcal{N}(0.315,\,0.014^{2})$
& $\mathcal{N}(0.315,\,0.014^{2})$ \\[6pt]

$\sigma_{\ln M}$
& $0.185$
& $0.185$
& $0.37$ \\[6pt]

\hline\hline
\end{tabular}
\label{tab:priors}
\end{table}

Cluster number counts are discrete events and therefore follow Poisson statistics. For each redshift and mass bin, $r,m$, the observed count
is $N_{r,m}^{\rm data}$, while the model predicts an expected mean count
$\mu_{rm}(\theta)$, which depends on the cosmological
parameters and nuisance parameters collected in $\theta$.
The Poisson probability of observing $N$ clusters given an expected mean
$\mu$ is
\begin{equation}
P(N \mid \mu) = \frac{\mu^N}{N!} e^{-\mu}.
\end{equation}

Assuming statistical independence between bins, the total likelihood is
the product of Poisson probabilities, and the log-likelihood becomes
\begin{equation}
\ln \mathcal{L}(\theta) =
\sum_{r,m} \left[
N^{\rm data}_{r,m}\,\ln \mu_{r,m}(\theta)
-
\mu_{r,m}(\theta)
- \ln\!\big(N^{\rm data}_{r,m}!\big)
\right].
\end{equation}
The final term is independent of $\theta$ and may be dropped in parameter inference. The expected counts in each bin are obtained via Eq.~\ref{eq: number counts 2}.

The log posterior is $\ln P(\theta|{\rm data})=\ln\mathcal{L}(\theta)+\ln p(\theta)$ where $p(\theta)$ denotes  the prior (see Table \ref{tab:priors}). It is sampled via Markov chain Monte Carlo (MCMC) with the affine-invariant ensemble sampler \texttt{emcee} \cite{foreman2013emcee}.

\section{RESULTS AND INTERPRETATION}
\label{sec:results}

\begin{figure}[t]
    \centering
\includegraphics[width=0.47\linewidth]{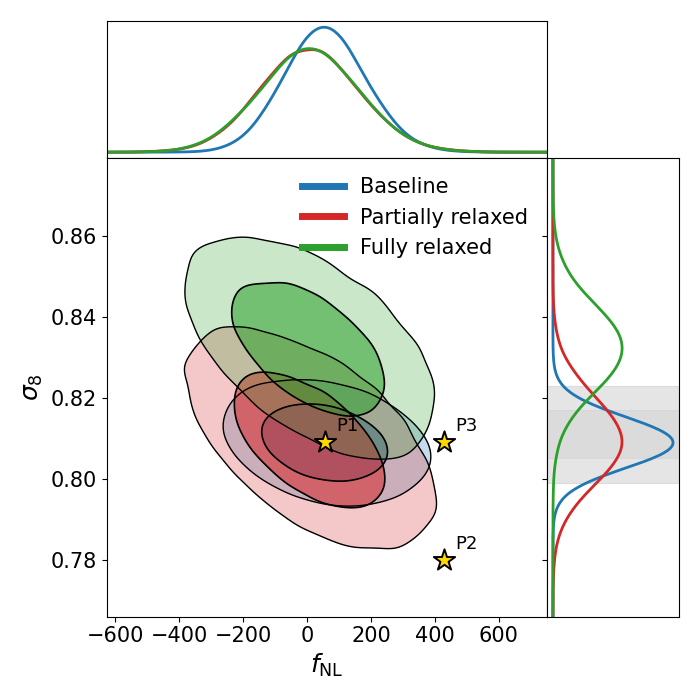}
\includegraphics[width=0.47\linewidth]{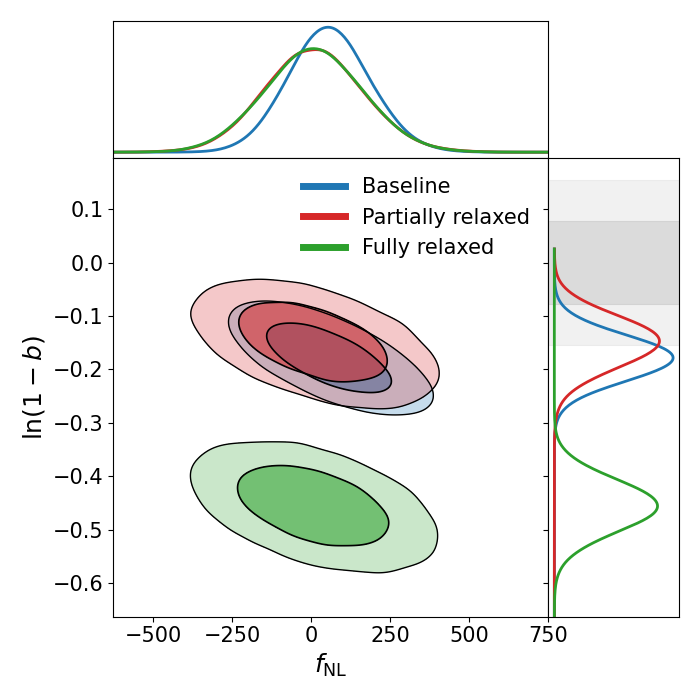}
\caption{Joint $f_{\rm NL}$–$\sigma_8$ (left) and $\ln(1-b)$–$f_{\rm NL}$ (right) posteriors for the three cases mentioned in Table~\ref{tab:priors}. Prior information for $\sigma_8$ and  $\ln(1-b)$ is shown as $1\sigma$ grey bands in the 1D marginals as discussed in the text. The $f_{\rm NL}$ posterior is robust to prior choices. The 3 points highlighted in the figure are used in later analysis and are as follows: P1 corresponds to the baseline best fit with $f_{\rm NL} = 55.3$, $\sigma_8 =0.809$ (blue curve in Fig.~\ref{fig:model-comparison-a}), P2 corresponds to $f_{\rm NL} = 430$, $\sigma_8 =0.78$ (orange curve in Fig.~\ref{fig:model-comparison-a}). P3 corresponds to $f_{\rm NL} = 430$, $\sigma_8 =0.809$ (green curve in Fig.~\ref{fig:model-comparison-a}). The other parameters remain fixed at the best fit values for the baseline case.  \label{fig:fNL_posterior}}
\end{figure}

\begin{figure}[h]
    \centering
\includegraphics[width=0.47\linewidth]{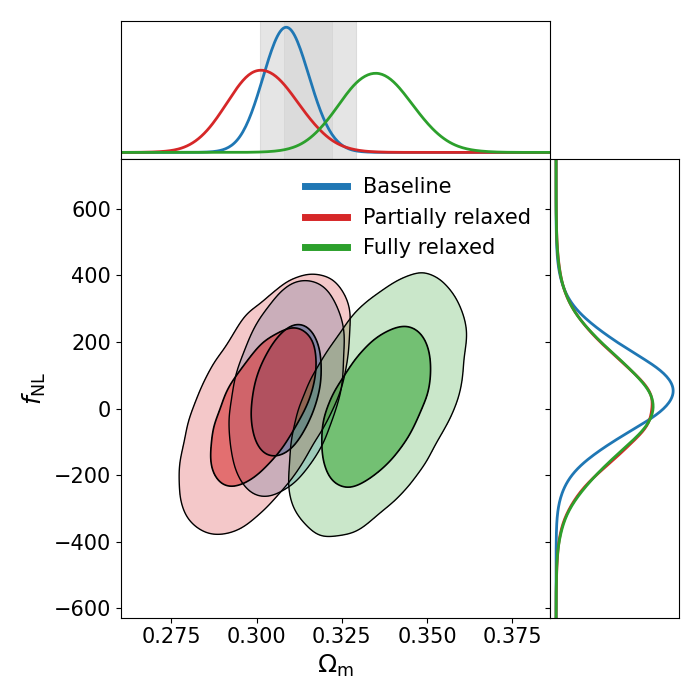}
\includegraphics[width=0.47\linewidth]{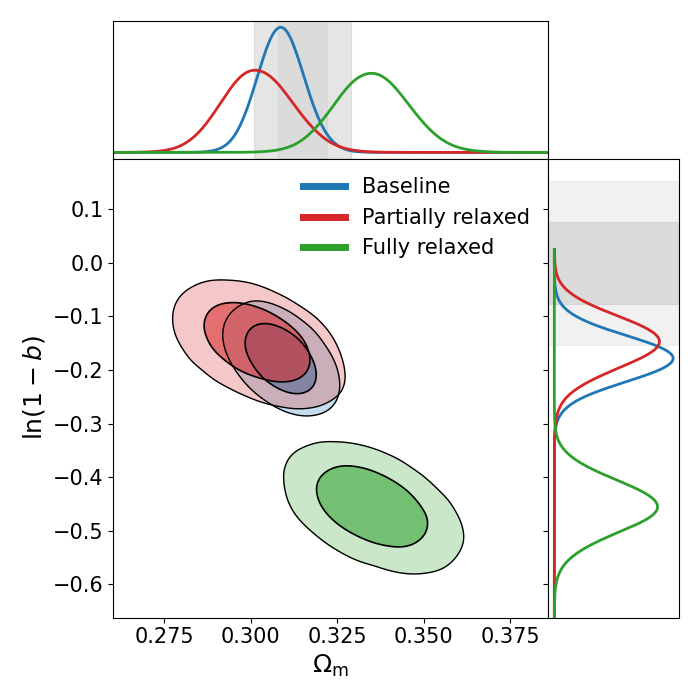}
\caption{Joint $f_{\rm NL}$–$\Omega_m$ (left) and $\ln(1-b)$–$\Omega_m$ (right) posteriors for the three cases mentioned in Table~\ref{tab:priors}. Prior information for $\Omega_m$ and  $\ln(1-b)$ is shown as $1\sigma$ grey bands in the 1D marginals, as discussed in the text.  While the $\Omega_m$ and mass bias posteriors overcome the priors,  the $f_{\rm NL}$ posterior is robust to prior choices.  \label{fig:Om_posterior}}
\end{figure}

Figs.~\ref{fig:fNL_posterior}--\ref{fig:Om_posterior} show the joint posteriors  for the baseline, partially relaxed and fully relaxed priors. The  relevant priors  are indicated as grey bands in the marginal distributions. The simultaneous display of the three cases highlights that relaxing the priors has a very mild effect on the $f_{\rm NL}$ constraints, although changes on the other parameters can be large.

\begin{figure*}[t]
    \centering

    % --------- Figure (a) ----------
    \begin{subfigure}[t]{\textwidth}
        \centering
        \includegraphics[width=\textwidth]{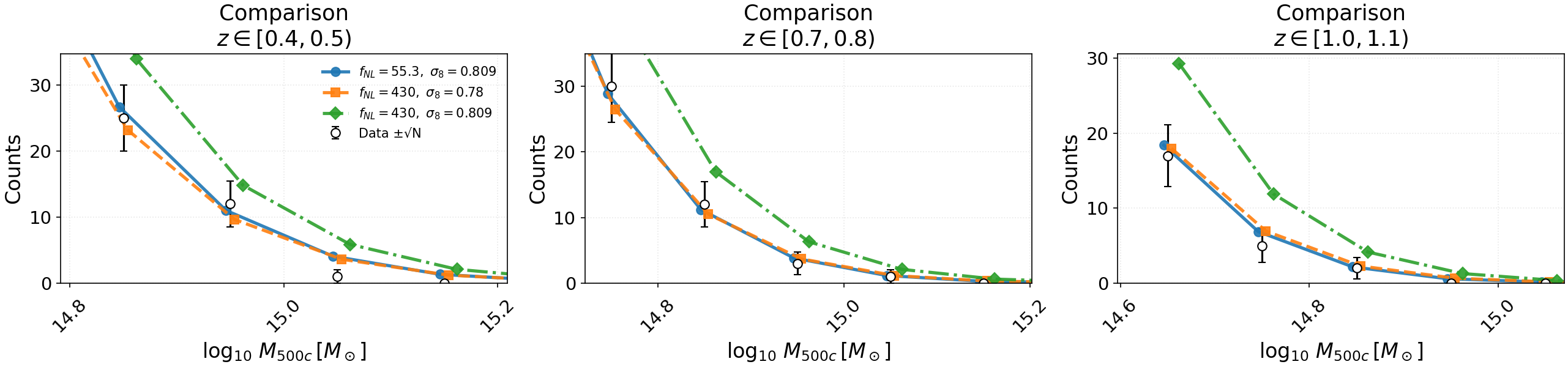}
        \caption{Cluster counts comparison.}
        \label{fig:model-comparison-a}
    \end{subfigure}

    \vspace{2em} % vertical spacing between figures

    % --------- Figure (b) ----------
    \begin{subfigure}[t]{\textwidth}
        \centering
        \includegraphics[width=\textwidth]{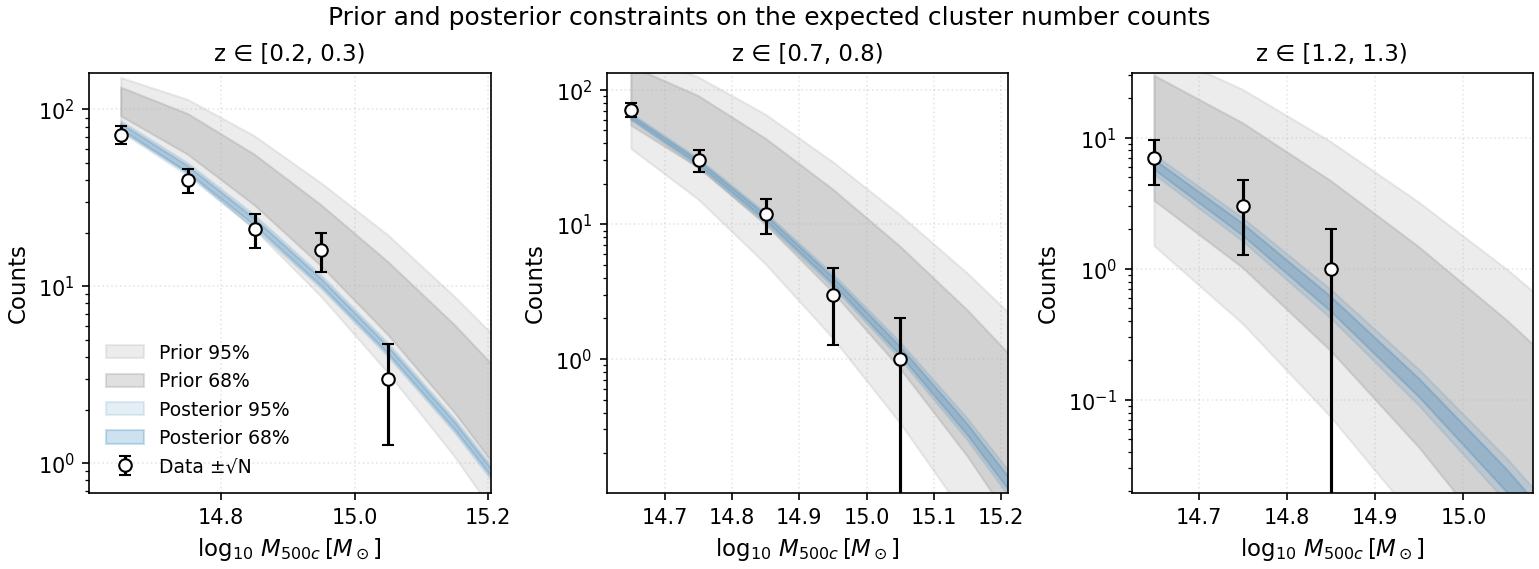}
        \caption{Constraints on the expected cluster number counts}
        \label{fig:model-comparison-b}
    \end{subfigure}

    \caption{Figure (a) shows a model comparison of cluster counts across redshift bins.
    Each panel shows the binned number of objects (Counts) as a function of
    $\log_{10} M_{500c}\,[M_\odot]$ for a different redshift interval:
    $z\in[0.4,0.5)$ (left), $z\in[0.7,0.8)$ (middle), and $z\in[1.0,1.1)$ (right).
    Open white circles with error bars represent the measured counts (shown as
    Data $\pm\sqrt{N}$). Colored curves show model predictions for different
    parameter choices. The comparison highlights the degeneracy between
    $\sigma_8$ and $f_{\mathrm{NL}}$, where distinct parameter combinations
    produce nearly identical number counts across all redshift bins. Note also that the $f_{\rm NL} =0$ plot would be very close to the blue/orange curves. Figure (b) 
    shows the prior and posterior bands on the expected cluster number counts. 
    This shows that the data is informative and constrains the number counts 
    into a thin strip in the counts-mass plane for each redshift bin.}
    \label{fig:model-comparison}
\end{figure*}

In particular the left panel of  Fig.~\ref{fig:fNL_posterior} shows the $\sigma_8-f_{\rm NL}$ posterior; relaxing the mass bias and mass scatter priors  affects the $\sigma_8$ posterior, leaving the $f_{\rm NL}$ constraint largely unaffected.
The stars indicate 3 points in parameter spaced  used in later analysis. P1 corresponds to the baseline best fit with $f_{\rm NL} = 55.3$, $\sigma_8 =0.809$ (blue curve in Fig.~\ref{fig:model-comparison-a}). For P2 and P3 we vary only the values of $f_{\rm NL}$ and $\sigma_8$:  P2 corresponds to $f_{\rm NL} = 430$, $\sigma_8 =0.78$ (orange curve in Fig.~\ref{fig:model-comparison-a}); P3 corresponds to $f_{\rm NL} = 430$, $\sigma_8 =0.809$ (green curve in Fig.~\ref{fig:model-comparison-a}).
It is interesting to note (right panel) that the posterior for the residual mass bias overcomes the prior (centered around $\ln(1-b)=0$) and prefers a significant negative residual mass bias: the shift is even more prominent for the fully relaxed prior case. This detection of a negative residual mass bias indicates that our analysis favors a residual mass bias in addition to the one that ACT incorporates into the inferred masses \cite{aguena2025atacama, robertson2024act}. The median of the residual mass bias posterior in the baseline case is reported to be (see Table \ref{tab:posteriors}) $\ln(1-b) = -0.179$ which implies $b=0.164$ i.e., a $16.4\%$ residual mass bias. This value is roughly $2.3 \sigma$ away from the prior mean, so it represents mild tension with the nominal WL calibration.

Fig.~\ref{fig:Om_posterior} shows the behavior of the $f_{\rm NL}$ posterior with $\Omega_m$ (left panel) and the effect of the residual mass bias on the recovered $\Omega_m$ (right panel). Note how the $\Omega_m$ posterior overcomes the prior and is driven  by the residual mass bias, while the $f_{\rm NL}$ constraint remains stable.  

The best fit values (posterior median) for  the three setups are reported in Table \ref{tab:posteriors}.
Despite  evidence of some correlation between $f_{\rm NL}$ and $\sigma_8$ and/or $\Omega_m$ the figures show that a much tighter external prior in any of these two parameters would not reduce significantly the uncertainty in $f_{\rm NL}$. We additionally tested the robustness of our results by varying the width of the mass and redshift and found no significant deviations from the main conclusions. 
\begin{table}[htbp]
\centering
\caption{Posterior constraints for the three analysis cases. Reported values are the posterior median with $1\sigma$ credible uncertainties.}
\begin{tabular}{lccc}
\hline\hline
Parameter 
& Baseline 
& Partially--relaxed 
& Fully--relaxed \\[2pt]
\hline

$f_{\rm NL}$ 
& $55.3 \pm 124.9$
& $7.9 \pm 152.3$
& $7.9 \pm 152.9$ \\[6pt]

$\ln(1-b)$ 
& $-0.179 \pm 0.040$
& $-0.149 \pm 0.046$
& $-0.456 \pm 0.047$ \\[6pt]

$h$
& $0.6756 \pm 0.0049$
& $0.6781 \pm 0.0098$
& $0.6720 \pm 0.0098$ \\[6pt]

$\sigma_{8}$
& $0.809 \pm 0.0056$
& $0.81 \pm 0.0106$
& $0.832 \pm 0.0105$ \\[6pt]

$\Omega_{m}$
& $0.309 \pm 0.0060$
& $0.302 \pm 0.0097$
& $0.3350 \pm 0.0101$ \\[6pt]

\hline\hline
\end{tabular}
\label{tab:posteriors}
\end{table}

To illustrate where the signal for the $f_{\rm NL}$ constraints comes from,  Fig.~\ref{fig:model-comparison-a} shows the measured number counts for selected redshift bins (points with error bars) and the prediction  for the best fit model (blue line) given by the baseline case from Table \ref{tab:priors}. The orange dashed curve shows number counts predictions for an alternative model which has an $f_{\rm NL}$ value $3\sigma$ away from the best fit value, with an alternative $\sigma_8$ chosen in such a way that they produce the same number counts as the baseline case, with all other parameters being the same in both models. This plot explicitly illustrates the $\sigma_8-f_{\rm NL}$ degeneracy. The tight prior on $\sigma_8$ partially breaks this degeneracy, informing the  inferred best-fit value for $f_{\rm NL}$ and other parameters. The green curve is an alternative model illustrating how a change in $f_{\rm NL}$ (while keeping all other parameters fixed to their baseline best-fit values) changes the number counts. All three models are also plotted as points in the joint $f_{\rm NL}-\sigma_8$ posterior graph in Fig.~\ref{fig:fNL_posterior}. Additionally varying the residual mass bias (along with $\Omega_m$) opens up  more extensive degeneracies in parameter space. 
Interestingly, even in the fully relaxed prior case, while $\sigma_8$, mass bias  and $\Omega_m$ shift significantly from the baseline best fit values, $f_{\rm NL}$ does not and remains stable. 

Further insights can be gained by comparing the prior-only prediction for the number counts with the posteriors and the data.
In Figure~\ref{fig:model-comparison-b},
the shaded gray regions indicate the $1\sigma$ and $2\sigma$ prior predictive bands, while the posterior predictive distribution collapses these bands into a narrow strip in the mass–number–count plane, demonstrating that the data is highly informative. This result should be interpreted together with Fig.~\ref{fig:model-comparison-a}. While Fig.~\ref{fig:model-comparison-b} shows that the data tightly constrain the mass dependence of the number counts, Fig.~\ref{fig:model-comparison-a} reveals that specific combinations of model parameters can reproduce essentially the same number counts across all redshifts. Consequently, the data constrain only those combinations of parameters rather than each parameter individually, and the final best-fit parameter values—shown in Figs.~\ref{fig:fNL_posterior}–\ref{fig:Om_posterior}—are therefore  informed by the priors. Yet, the recovered $f_{\rm NL}$ value appears robust to the prior choice, within the prior range explored here.

\clearpage

\section{CONCLUSIONS}
\label{sec:conclusions}

Using the ACT DR6 Sunyaev–Zeldovich–selected galaxy cluster catalog, we derived constraints on primordial non-Gaussianity of the local type by modeling cluster abundances with a Log–Edgeworth halo mass function and a forward-modeled likelihood that incorporates completeness, residual mass bias, intrinsic scatter, and $f_{\rm NL}$. Our analysis of 1,201 clusters over 10,347 square degrees of sky yields a posterior mean of $f_{\rm NL} = 55 \pm 125$ (68\% CL), consistent with a Gaussian initial density field at comoving cluster scales of 5 to 10 Mpc $h^{-1}$.
This constraint adopts Gaussian priors on cosmological parameters  $\sigma_8$, $\Omega_m$ and $h$ from Planck 2018 \cite{aghanim2020planck}, and motivated priors on the residual mass bias and mass scatter from \cite{aguena2025atacama} (see sec 4.1).
The scales probed by this measurement complement CMB bispectrum analyses, probing modes not directly accessible through Planck or other CMB experiments.

Our analysis indicate that any discrepancy between observed and Gaussian-predicted number counts is better explained by a significant residual mass bias than by a non-zero $f_{\rm NL}$, given the adopted priors. There is a strong degeneracy among parameters $f_{\rm NL}$, residual mass bias, and $\sigma_8$ (and of the mass bias with $\Omega_m$)  which is only lifted by the external priors. This finding underscores the importance of accurate mass calibration and bias priors: uncertainties in the SZ mass–observable relation dominate the error budget and can mimic PNG signals.

Although this constraint is less stringent than Planck’s $f_{\rm NL} = -0.9 \pm 5.1$, it provides an independent test probing much smaller scales and late times,   relying on different physics and using a different observable. The ACT SZ sample, approximately mass limited and nearly redshift independent, offers a robust basis for future multi-probe analyses. Upcoming wide-field optical, SZ, and lensing surveys (e.g., Rubin/LSST, Euclid, Simons observatory) with improved mass calibration are expected to tighten cluster-based PNG constraints significantly and enable further tests of possible systematics. While Planck CMB limits currently set the tightest bounds on simple local-type PNG, cluster abundances remain a valuable, independent probe — particularly for scale-dependent and high-order PNG signatures that some inflationary models predict.

\appendix
\section{EXPLICIT EXPRESSIONS OF HALO MASS FUNCTIONS}
\label{sec:appendix}
In this appendix, for completeness,  we review the halo mass functions that appear in the literature and also briefly quote the cluster count predictions obtained from one of them. We emphasize the point that these mass functions suffer from some drawbacks that are discussed below.

The two most widely used ones are the MVJ \cite{matarrese2000abundance} and LoVerde \cite{loverde2008effects} ones. We begin with the MVJ mass function:
\begin{equation}
n(M,z) \;=\; 
\frac{2\,\bar{\rho}_m\,\Omega_{m,0}\,H_0^2}{8\pi G M^2}\,
\frac{1}{\sqrt{2\pi}\,\sigma_M}\,
\exp\!\left[-\frac{\delta_*^2}{2\sigma_M^2}\right]\,
\left[
\frac{1}{6}\left(\frac{\delta_c^2}{\sigma_M^2}-1\right)\,
\frac{dS_{3,M}}{d\ln M}
+ \frac{\delta_*}{\sigma_M}\,\frac{d\sigma_M}{d\ln M}
\right].
\label{eq:matarrese2000}
\end{equation}
\noindent
The parameter $\delta_\ast$ is a shifted critical density threshold that incorporates the effect of the skewness of the smoothed density field. It is defined as $
\delta_\ast \;\equiv\; \delta_c \,\sqrt{\,1 - \tfrac{1}{3}\,\delta_c\,S_{3,M}}\,,$
where $\delta_c \simeq 1.686$ is the linear overdensity for spherical collapse, and $S_{3,M}$ is the normalized skewness of the density field smoothed on mass scale $M$. The normalized skewness is defined as $S_3(M)\equiv\langle\delta_M^3\rangle/\sigma^4(M)$.

An alternative approach was used by LoVerde et al.\ (LV) \cite{loverde2008effects} to derive similar results using an Edgeworth expansion:
 \begin{align}
n(M,z) \;=\;&\; \frac{3 H_0^2 \,\Omega_{m,0}}{8\pi G M^2}\;
\frac{1}{\sqrt{2\pi}\,\sigma_M}\;
\exp\!\left[-\frac{\delta_c^2}{2\sigma_M^2}\right] \nonumber \\[6pt]
&\times \left\{ 
\frac{d\ln\sigma_M}{dM}
\left(\frac{\delta_c}{\sigma_M} 
+ \frac{S_{3,M}\,\sigma_M}{6}
\left[\frac{\delta_c^4}{\sigma_M^4}
- 2\,\frac{\delta_c^2}{\sigma_M^2} - 1 \right]\right)
+ \frac{1}{6}\,\frac{dS_{3,M}}{dM}\,\sigma_M
\left(\frac{\delta_c^2}{\sigma_M^2} - 1\right)
\right\}.
\label{Lo Verde non-Gaussian hmf}
\end{align}

These analytic prescriptions agree qualitatively with simulations but require calibration. Grossi et al. \cite{Grossi2009} showed that excellent agreement with $N$-body results is obtained by rescaling the collapse threshold, $\delta_c\rightarrow \delta_c\sqrt{q}$ with $q\simeq 0.75$. They proposed that the full non-Gaussian halo mass function should be written as
\begin{equation}
n_{\mathrm{NG}}(M,z,f_{\mathrm{NL}}) \;=\;
n^{\mathrm{sim}}_{G}(M,z) \;\times\; R_{\mathrm{NG}}(M,z)
\label{non-Gaussian-hmf}
\end{equation}
where $n^{\mathrm{sim}}_{G}(M,z)$ is the Gaussian halo mass function calibrated to N-body simulations (e.g. Tinker or ST), meanwhile $R_{\mathrm{NG}}(M,z)$ is the non-Gaussian multiplicative factor and, for the Lo Verde halo mass function , is given by 
\begin{equation}
R_{\mathrm{NG}}(M,z,f_{\mathrm{NL}}) \;=\;
1 + \frac{1}{6}\,\frac{\sigma_M^2}{\delta_{ec}}
\left[
S_{3,M}\!\left(\frac{\delta_{ec}^4}{\sigma_M^4}
- 2\frac{\delta_{ec}^2}{\sigma_M^2} - 1 \right)
+ \frac{dS_{3,M}}{d\ln\sigma_M}
\left(\frac{\delta_{ec}^2}{\sigma_M^2} - 1\right)
\right]\,,
\label{R factor}
\end{equation}
with the rescaled $\delta_{ec} = \delta_c \sqrt{q}$ for $q=0.75$. Path-integral and excursion-set approaches \cite{MaggioreRiotto2010a,MaggioreRiotto2010b,DAmico2011} provide alternative routes as well as further analytic insights. Recent simulation-based studies continue to refine non-Gaussian halo mass function calibrations, emphasizing the dependence on halo definition and overdensity threshold and proposing practical reparameterizations of $\delta_c$ to avoid parameter biases in $f_{\rm NL}$ inference (e.g., \cite{Fiorino2025}).

Both the plain Edgeworth expansion of LV and the MVJ approaches have two notable drawbacks: they are truncated at low order in the cumulants, so they cannot systematically include higher-order effects such as kurtosis or variance corrections, and more critically, they can become negative (or imaginary in MVJ's case) in the rare-event tail,
producing unphysical probabilities for very massive haloes. Hence in the main text we use the Log-Edgworth formulation that avoids these issues. 

\begin{figure}[t]
  \centering
  \begin{minipage}[t]{0.49\linewidth}
    \centering
    \includegraphics[height=4.2cm]{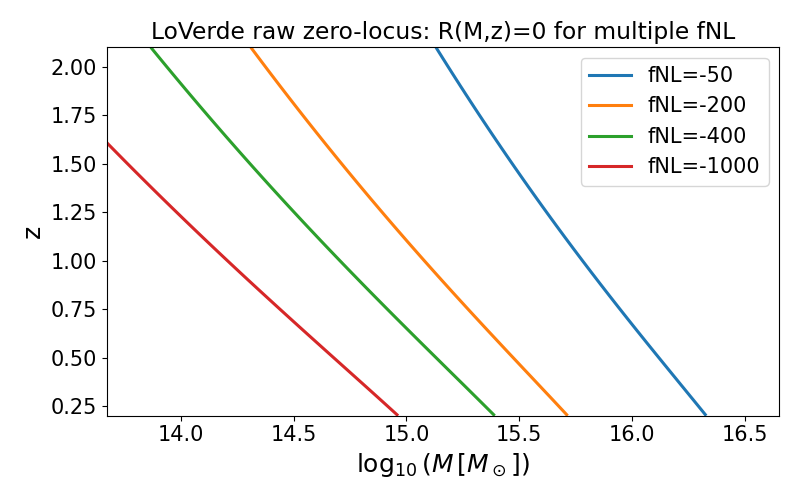}
    \subcaption{$R(M,z)=0$ in the $(\log_{10}M,z)$ plane for several negative $f_{\rm NL}$.}
    \label{fig:zero_locus}
  \end{minipage}\hfill
  \begin{minipage}[t]{0.49\linewidth}
    \centering
    \includegraphics[height=4.2cm]{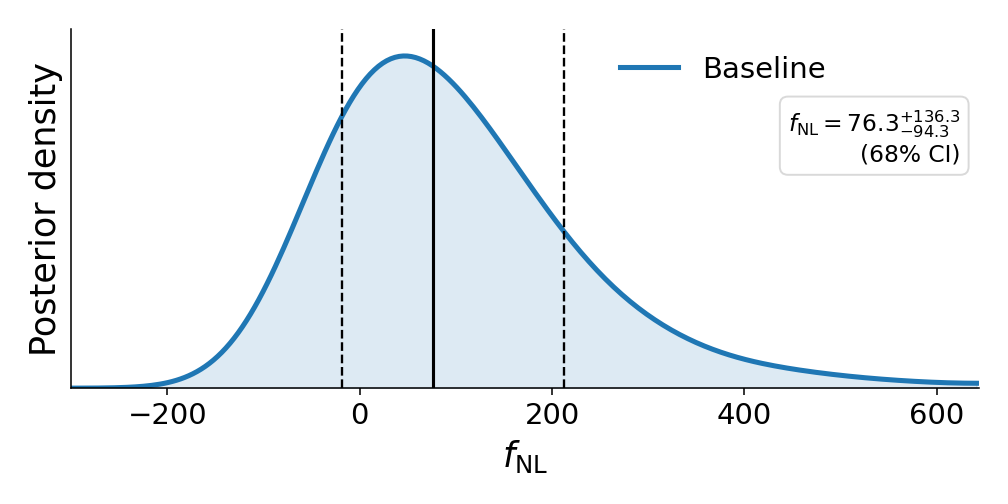}
    \subcaption{Posterior of $f_{\rm NL}$ (solid: median; dashed: 68\% CI).}
    \label{fig:loverde_hmf_fnl_posterior}
  \end{minipage}
  \caption{Diagnostics for the LoVerde mass function analysis.}
  \label{fig:loverde_diagnostics}
\end{figure}

We use the LV mass function to illustrate these drawbacks and how they affect results. Fig.~\ref{fig:loverde_hmf_fnl_posterior} plots the $f_{\rm NL}$ posterior, and Table \ref{tab:LV posteriors_baseline} quotes the values of all the parameters in the baseline case. An important point is that the posterior is noticeably asymmetric and is shifted toward positive values. This behavior arises because at certain negative values of $f_{\rm NL}$, Eq.~\ref{R factor} can become negative, signaling the breakdown of the perturbative expansion.

In practice, such parameter values are non-physical and are therefore rejected during MCMC sampling. This effectively introduces a lower cutoff in the $f_{\rm NL}$ posterior, beyond which negative values cannot be explored. Importantly, this cutoff is not universal: it depends on both halo mass and redshift, as is evident from Eq.~\ref{R factor}. For clarity, we also show the locus where $R=0$ in Fig.~\ref{fig:zero_locus}. As can be seen, this condition is reached over a substantial region of the data set for moderate to large  and  negative $f_{\rm NL}$ values, indicating that the perturbative description is violated in a non-negligible portion of the parameter space and that the resulting constraints on $f_{\rm NL}$ should therefore be interpreted with caution. 

\begin{table}[htbp]
\centering
\caption{Posterior constraints for the baseline analysis for the LV mass function. Reported values are the posterior median with $68\%$ credible intervals. 
The constraint on $f_{\rm NL}$ is explicitly asymmetric.}
\begin{tabular}{lc}
\hline\hline
Parameter & Baseline \\[4pt]
\hline

$f_{\rm NL}$ 
& $76.3^{+136.3}_{-94.3}$ \\[6pt]

$\ln(1-b)$
& $-0.197 \pm 0.051$ \\[6pt]

$h$
& $0.6757 \pm 0.0049$ \\[6pt]

$\sigma_{8}$
& $0.8080 \pm 0.0059$ \\[6pt]

$\Omega_{m}$
& $0.3090 \pm 0.0058$ \\[6pt]

\hline\hline
\end{tabular}
\label{tab:LV posteriors_baseline}
\end{table}

\clearpage

\section*{ACKNOWLEDGEMENTS} 
We thank Boris Bolliet, Andrina Nicola, Eunsong Lee, and Nick Battaglia for their work on cluster mass determination and catalog systematics. 
LS, RJ and LV acknowledge support of Spanish MINECO under project  PID2022-141125NB-I00 MCIU/AEI,
and the ``Center of Excellence Maria de Maeztu 2025-2029'' award to the ICCUB (CEX2024-001451-M funded by MICIU/AEI/10.13039/501100011033).
MH acknowledges support from the National Research Foundation of South Africa (grant nos. 97792, 137975, CPRR240513218388).
Support for ACT was through the U.S.~National Science Foundation through awards AST-0408698, AST-0965625, and AST-1440226 for the ACT project, as well as awards PHY-0355328, PHY-0855887 and PHY-1214379. Funding was also provided by Princeton University, the University of Pennsylvania, and a Canada Foundation for Innovation (CFI) award to UBC. ACT operated in the Parque Astron\'omico Atacama in northern Chile under the auspices of the Agencia Nacional de Investigaci\'on y Desarrollo (ANID). The development of multichroic detectors and lenses was supported by NASA grants NNX13AE56G and NNX14AB58G. Detector research at NIST was supported by the NIST Innovations in Measurement Science program. Computing for ACT was performed using the Princeton Research Computing resources at Princeton University, the National Energy Research Scientific Computing Center (NERSC), and the Niagara supercomputer at the SciNet HPC Consortium. SciNet is funded by the CFI under the auspices of Compute Canada, the Government of Ontario, the Ontario Research Fund–Research Excellence, and the University of Toronto. We thank the Republic of Chile and the local indigenous Licanantay communities for hosting ACT in the northern Atacama.
\bibliographystyle{JHEP}
\bibliography{references}
\end{document}